\documentclass[graybox]{svmult}

\usepackage{mathptmx}       % selects Times Roman as basic font
\usepackage{helvet}         % selects Helvetica as sans-serif font
\usepackage{courier}        % selects Courier as typewriter font
\usepackage{makeidx}         % allows index generation
\usepackage{graphicx}        % standard LaTeX graphics tool
\usepackage{multicol}        % used for the two-column index
\usepackage[bottom]{footmisc}% places footnotes at page bottom
\usepackage{subfigure}
\usepackage{subeqnarray}
\usepackage{amsmath}
\usepackage{amssymb}

\makeindex             % used for the subject index
\newcommand{\bmu}{\mathbf{u}}

\newcommand{\pp}{\partial}

\begin{document}

\title*{A fast direct solver for the advection-diffusion equation using low-rank approximation of the Green's function}
\titlerunning{Fast direct solver for advection-diffusion equation}
\author{Jonathan R. Bull}
\authorrunning{J. R. Bull}
\institute{Division of Scientific Computing, Uppsala University, Polacksbacken, Uppsala, \email{jonathan.bull@it.uu.se}}

\maketitle

\abstract*{}

\abstract{
We present a fast direct solution method for the advection-diffusion equation in one and two dimensions with non-periodic boundaries.
Computational cost is reduced to $\mathcal O(N)$ by making a low-rank approximation of the Green's function without sacrificing accuracy.
Implicit treatment of the diffusion term reduces stiffness in advection-dominated problems.
Results show that the solver is roughly an order of magnitude faster than a reference method, namely the Matlab backslash operator.
This work motivates the use of hierarchical low-rank approximations for solution of stiff hyperbolic problems at very large scale, including those arising from high-order accurate spatial discretisations.
}

%\begin{svgraybox}
%If you want to emphasize complete paragraphs of texts we recommend to use the newly defined Springer class option \verb|graybox| and the newly defined environment \verb|svgraybox|. This will produce a 15 percent screened box 'behind' your text.
%\end{svgraybox}

%%%%%%%%%%%%%%%%%%%%%%%%%%%%%%%%%%%%%%%%%

\section{Introduction}

%background
The latest supercomputers derive their processing power from executing millions of threads simultaneously across thousands of processors.
Unfortunately, the solution algorithms at the heart of computational physics codes are not well suited to operating in this way and linear scaling is generally lost.
As an example, Newton-Krylov (NK) solvers require a communication-intensive global inner product calculation every iteration \cite{mcinnes:2014}.
%main topic/previous work
This is a strong motivation to develop new solution algorithms that can achieve better scaling in the new HPC environment by reducing communication and increasing algebraic intensity.
Hierarchical algorithms such as the Fast Multipole Method (FMM) \cite{greengard:87} and $\mathcal H^2$ matrices \cite{hackbusch:2000} have considerable potential in this arena.
They are increasingly finding application as direct solvers and low-cost preconditioners for certain PDEs and boundary integral equations \cite{keyes2013,darve:14,yokota:16}.
Elliptic and parabolic PDEs can be solved by hierarchical methods in $\mathcal O(N \log N)$ or $\mathcal O(N)$ operations (where $N$ is the number of degrees of freedom) and may also outperform multigrid (MG) \cite{keyes2013,gholami:14}.

Hierarchical methods are only useful for PDEs with a Green's function that decays rapidly enough between two points in space, such as the Laplace and Poisson equations.
In that case a hierarchical low-rank approximation (HRLA) of the inverse of the discrete operator can be computed by hierarchically compressing off-diagonal blocks.
It can be shown that the approximation satisfies an error bound, as done by \cite{engquist:2011} for the Helmholtz equation.
For an excellent review of HRLA see \cite{yokota:16}.

%new ideas
HRLA cannot be applied directly to hyperbolic PDEs because their Green's functions do not have the necessary decaying property.
However, Bull et al. \cite{bull:16b} showed that the semi-discrete advection-diffusion equation could be transformed into a parabolic form that could be solved by HRLA.
By discretising the 1D periodic advection-diffusion equation in time such that the advective term is explicit and the diffusion term implicit, the linear system became parabolic.
A simple non-hierarchical low-rank approximation of the Green's function was used as a proof of concept.
Results obtained in 1D with periodic boundaries demonstrated linear scaling of CPU time with $N$.

%paper outline
The ideas first presented in \cite{bull:16b} are further developed here: the derivation is put on a more analytical footing and the method is extended to the 2D advection-diffusion equation and to non-periodic domains.
The paper has six parts: in \S 2 the 1D solver is presented; in \S 3 the 2D solver is presented including the method of images for imposing Dirichlet/Neumann boundary conditions; in \S 4 the low-rank approximation schemes are described; in \S 5 and \S 6 the 1D and 2D numerical tests are presented; in \S 7 conclusions are drawn and future work proposed.

%%%%%%%%%%%%%%%%%%%%%%%%%%%%%%%%%%%%%%%%%

\section{Direct solver in 1D}\label{1d}

\subsection{1D periodic advection-diffusion equation}

We consider the advection-diffusion equation in 1D:
\begin{equation}\label{advdiff}
  \frac{\pp u}{\pp t} + a\frac{\pp u}{\pp x} - \nu \frac{\pp^2 u}{\pp x^2} = 0,
\end{equation}
where $u$ is a smooth scalar field, $a$ is the advection coefficient (linear or nonlinear), $\nu$ is the diffusion coefficient (constant), and periodic boundary conditions are imposed.
The central idea is to transform the equation into a parabolic equation such that it can be solved using fast direct methods.
We start by putting the advection term on the right-hand side (RHS):
\begin{equation}\label{advdiff2}
  \frac{\pp u}{\pp t} - \nu \frac{\pp^2 u}{\pp x^2} = -a\frac{\pp u}{\pp x},
\end{equation}
The conventional approach to solving this equation is to discretise in time and space to obtain
%Now, the semi-discrete equation is written at a timestep $t^{n+1}$:
%
%\begin{subeqnarray}\label{advdiff3}
%  \frac{u^{n+1}-u^{n}}{\Delta t} - \nu \frac{\pp^2 u}{\pp x^2} = -a\frac{\pp u}{\pp x}, \\
%  \therefore u^{n+1} - \nu \Delta t \frac{\pp^2}{\pp x^2}(u^{n+1}) = u^{n} -a\Delta t \frac{\pp u^n}{\pp x},
%\end{subeqnarray}
%
%where the diffusion term has been made implicit in time and the advection term is explicit.
%This is the simplest IMEX scheme; higher-order IMEX schemes can also be applied such as those in \cite{ascher:95}.
%The equation now resembles the diffusion or heat equation with a source term.
%Discretising in space we obtain
%
\begin{equation}\label{advdiff4}
  (I - \Delta t D) \bmu^{n+1} = (I - \Delta t A) \bmu^n,
\end{equation}
and solve the linear system iteratively.
However, the continuous equation \eqref{advdiff2} can be solved directly at time $t^{n+1}$ by a convolution with the Green's function over the domain and a time interval $[t^n, t^{n+1}]$:
\begin{eqnarray}\label{green1}
  u^{n+1}(x) &= \int_{t^{n}}^{t^{n+1}} \int_{-\infty}^{\infty} G(x,x',t^{n+1},t') f(x',t') dx' dt' \nonumber \\
  &+ \int_{-\infty}^{\infty} G(x,x',t^{n+1},t^{n}) u^n(x') dx', \\
  f(x,t) &= -a \frac{\partial}{\partial x} (u(x,t)), \\
  G(x,x',t,t') &= (4\pi \nu(t-t'))^{-1/2} \exp \left ( {-\frac{|x-x'|^2}{4\nu(t-t')}} \right ). 
\end{eqnarray}
This method is similar to exponential integrators for solving linear ODEs, except that this is an `analytical implicit' method while exponential integrators are `analytical explicit' methods.
We have considerable freedom in choosing approximations for the space and time integrals in \eqref{green1}.
A high-order approximation could be used for time integral in \eqref{green1} but for simplicity the Green's function is linearised about $t^n$ and the forcing function is evaluated at $t^n$ (explicit Euler approximation):
\begin{equation}\label{green2}
  u^{n+1}(x) = \int_{-\infty}^{\infty} G(x,x',t^{n+1},t^{n}) [ u^n(x') + \Delta t f(x',t^n) ] dx'.
\end{equation}
Defining a wavenumber $\epsilon = (4\nu \Delta t)^{-1/2}$ and distance $r=|x-x'|$:
\begin{equation}\label{green3}
  u^{n+1}(x) = \int_{-\infty}^{\infty} \frac{\epsilon}{\sqrt{\pi}} \exp({-\epsilon^2 r^2}) [u^n(r) + \Delta t f(u^n(r))] dr.
\end{equation}
Now the problem is restricted to a finite periodic domain of length $L$ with $N$ intervals of uniform size $\Delta x = L/N$.
Piecewise-constant approximation of the space integral in \eqref{green3} with an upwind finite difference discretisation of the advection term leads to
\begin{eqnarray}\label{green4}
  &u_i^{n+1} = M_{ij} (u_j^n+\Delta t f_j^n), \quad i = 1,\ldots,N, \\
  &= \sum_{j=i-N/2}^{i+N/2} \frac{\epsilon \Delta x}{\sqrt{\pi}} \exp(-(|j-i| \epsilon \Delta x)^2) [(1-c_A) u^n_j +c_A u^n_{j-1}],
\end{eqnarray}
where the advective CFL number $c_A = a\Delta t/\Delta x$.
Note the equivalence of terms in \eqref{green4} to those in \eqref{advdiff4}: the RHS term $(u_j^n+\Delta t f_j^n) \equiv (I - \Delta t A) \bmu^n$ and the Green's matrix $M \equiv (I - \Delta t D)^{-1}$.

\begin{figure}[hbt]
  \subfigure[periodic]{
  \includegraphics[width=0.48\textwidth]{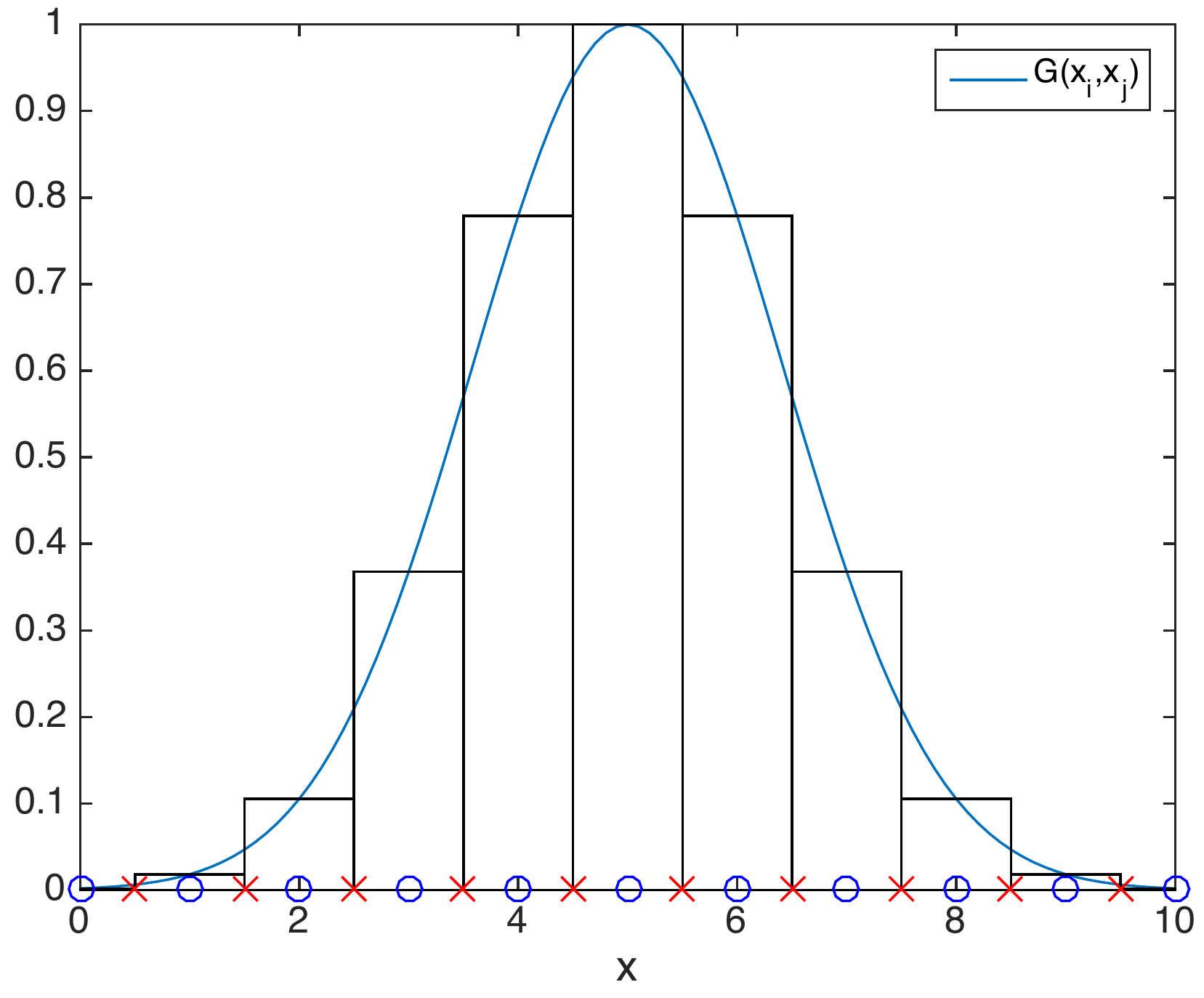}}
  \subfigure[non-periodic]{
  \includegraphics[width=0.48\textwidth]{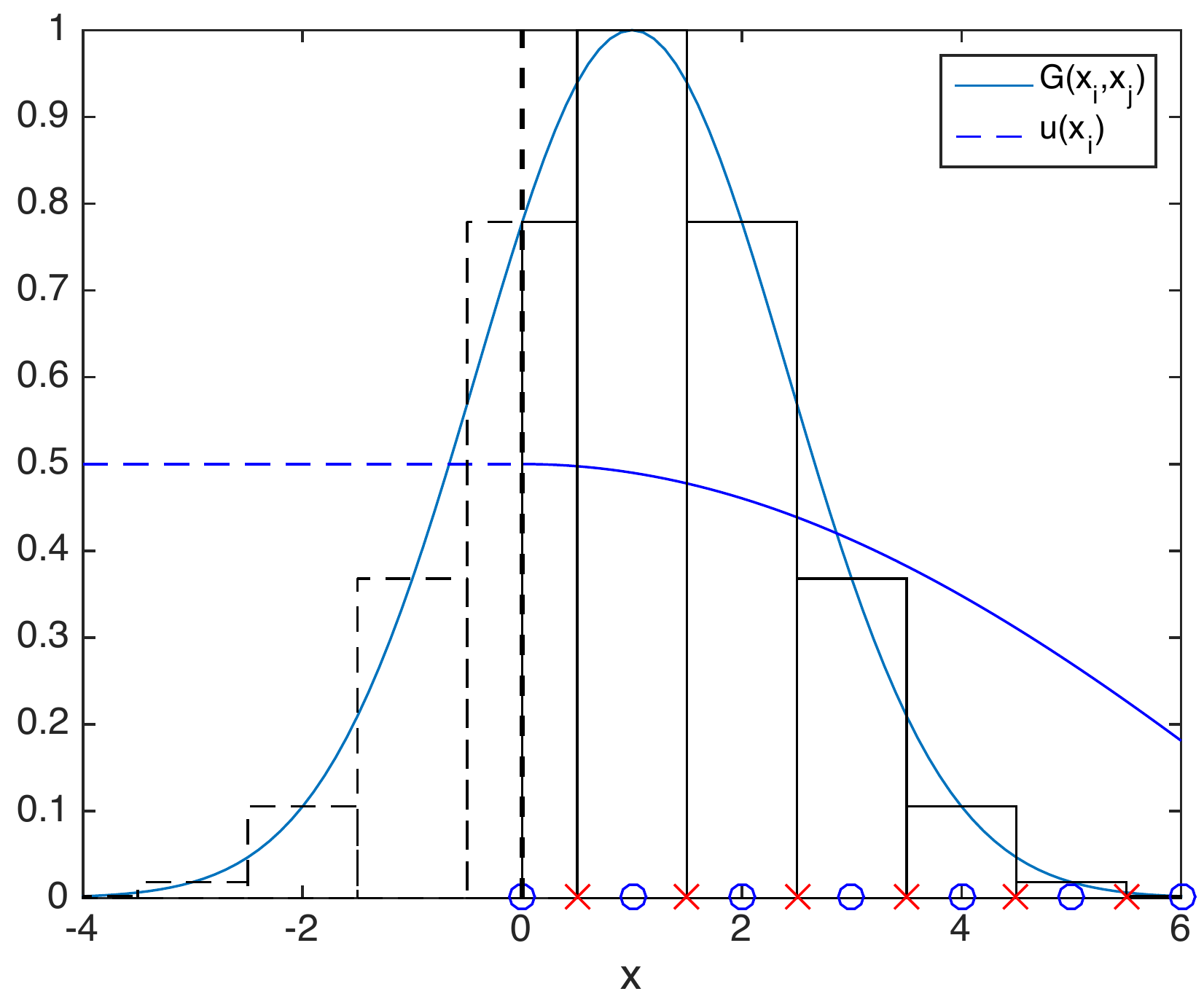}}
  \caption{Piecewise-constant approximation (black) of the Green's function (light blue). Circles indicate the nodes $x_i$, crosses denote the interval midpoints.
  In (b), vertical dashed line is the boundary and the solution is extended outside domain (dark blue dashes) and multiplied by the exterior portion of Green's function.}
  \label{greenfig}
\end{figure}

Periodicity is imposed on the index $j$ such that if $j-i>N/2$ then $j=j-N$ and similarly, if $j-i<N/2$ then $j=j+N$.
The piecewise-constant integral approximation is illustrated in Figure \ref{greenfig}.
The $i$th interval lies in the range $x \in [x_{i-1/2},x_{i+1/2}]$ and has value $G(x_i)$.
Hence the first and $N$th intervals are bisected by the domain boundaries.
Due to periodicity they are two halves of the same interval.

The periodic advection-diffusion equation is solved by a single matrix-vector product per timestep with computational cost of $\mathcal O(N^2)$.
If the timestep is fixed and the diffusion coefficient is constant in time then the matrix $M_{ij}$ can be assembled just once and used repeatedly; only the RHS changes each time iteration.
The method also extends easily to non-uniform grids and to nonlinear advection.
One can also use finite volume or finite element schemes for approximating \eqref{green3}.

\subsection{1D advection-diffusion equation with Dirichlet/Neumann boundary conditions}

To impose a Dirichlet boundary condition on the inflow (let's say the left boundary), $g_L = u(x=0) = u_1$, the solution is extended by $g_L$ beyond the left boundary to $x=-L$.
To impose a zero Neumann BC on the outflow (right boundary), the solution is extended by $g_R=u(x=L) = u_N$ beyond the right boundary to $x=2L$.
Then, the problem is solved by
\begin{eqnarray}\label{greenbc}
  u_i^{n+1} &= M_{ij} (u_j^n+\Delta t f_j^n) + B^L_i g_L + B^R_i g_R \nonumber \\
  &= \sum_{j=1}^{N} \frac{\epsilon \Delta x}{\sqrt{\pi}} \exp(-((j-i) \epsilon \Delta x)^2) [(1-c_A) u^n_j +c_A u^n_{j-1}] \nonumber \\
  &+ \sum_{j=\min(i+1,N)}^{N} \frac{\epsilon \Delta x}{\sqrt{\pi}} \exp(-(j-1) \epsilon \Delta x)^2) g_L \nonumber \\
  &+ \sum_{j=1}^{\max(1,i-1)} \frac{\epsilon \Delta x}{\sqrt{\pi}} \exp(-((j-N) \epsilon \Delta x)^2) g_R.
\end{eqnarray}
The two `boundary influence vectors' $B^L$ and $B^R$ are simply summations of the part of the discrete Green's function centred on a point $i$ that lies outside the respective boundary, as illustrated in Figure \ref{greenfig} (b) for the left boundary.
That is, the domain of integration of the $i$th point is extended from $[0,L]$ to $[\min(0,x_i-L/2), \max(L,x_i+L/2)]$.
This method of imposing boundary conditions is essentially the method of images \cite{olver:13}.
To save effort the boundary vectors can be computed beforehand and stored, as with the matrix M.
Because $M$ is circulant, $B^L$ and $B^R$ can be calculated from the first row:
\begin{subeqnarray}
	B^L_i &= \sum_{j=i+1}^{N/2} M_{1,j}, \ i=1,\ldots,N, \\
	B^R_i &= B^L_{N-i+1}.
\end{subeqnarray}
%

%%%%%%%%%%%%%%%%%%%%%%%%%%%%%%%%%%%%%%%%%

\section{Direct solver in 2D}\label{2d}

The method above generalises easily to 2D structured grids.
Performing the same splitting of the continuous equation as before,
\begin{eqnarray}\label{green2d1}
  u^{n+1}(x,y) &= \int_{t^{n}}^{t^{n+1}} \int_{-\infty}^{\infty} \int_{-\infty}^{\infty} G(x,x',y,y',t^{n+1},t') f(x',y',t') dx' dy' dt' \nonumber \\
  &+ \int_{-\infty}^{\infty} \int_{-\infty}^{\infty} G(x,x',y,y',t^{n+1},t^{n}) u^n(x',y') dx' dy', \\
  f(x,y,t) &= u^n - a \cos(\theta) \frac{\partial}{\partial x}u(x,y,t) - a \sin(\theta) \frac{\partial}{\partial y}u(x,y,t), \\
  G(x,x',y,y',t,t') &= 4\pi \nu(t-t') \exp \left ( {-\frac{(x-x')^2+(y-y')^2}{4\nu(t-t')}} \right ).
\end{eqnarray}
Treating the time integral as before and applying finite differences on an $P \times P$ grid with $N = P^2$ total degrees of freedom:
\begin{subeqnarray}\label{green2d2}
  u_{ij}^{n+1} &=& M_{ijkl} (u_{kl}^n+\Delta t f_{kl}), \quad i,j = 1,\ldots,P, \\
  &=& \sum_{k=i-P/2}^{i+P/2} \ \sum_{l=j-P/2}^{j+P/2} \frac{\epsilon^2 \Delta x \Delta y}{\pi} \nonumber \\
  && \exp \left[-\epsilon^2 ((|k-i| \Delta x)^2 + (|l-j| \Delta y)^2) \right] (u_{kl}^n+\Delta t f_{kl}^n).
\end{subeqnarray}
The 2D periodic advection-diffusion equation is solved by \eqref{green2d2} with a single matrix-vector product per timestep.
Although quadruple indexing has been used here, the solver is implemented as a matrix-vector product of dimensions $[N,N]\times(N)$.
We refer to $u_{ij}$ as a vector and $M_{ijkl}$ as a dimension-two matrix.

%%%%%%%%%%%%%%%%%%%%%%%%%%%%%%%

\begin{figure}[hbt]
  \sidecaption
  \includegraphics[width=0.64\textwidth]{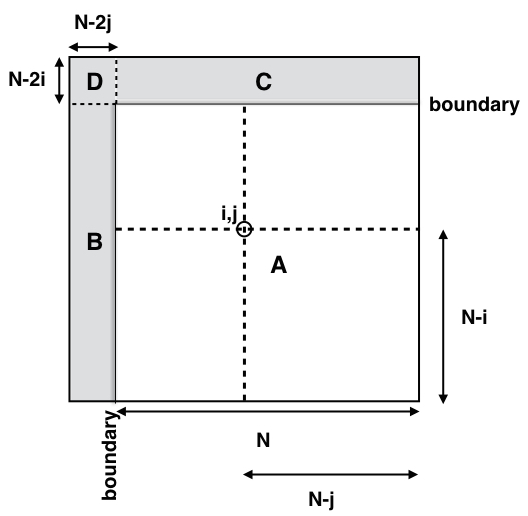}
  \caption{Method of discrete Green's function in 2D at a Dirichlet/Neumann boundary. Here $N$ refers to the number of points per dimension.}
  \label{greenbc2dfig}
\end{figure}

In the 1D non-periodic case, the Green's function lying beyond a boundary was integrated and multiplied by the Dirichlet value.
The same is true in the 2D non-periodic case but now we must integrate over an area.
A simplified form of a boundary integral method is employed, illustrated in Figure \ref{greenbc2dfig}.
Dirichlet conditions $g^L, g^T$ are imposed on the left and top boundaries representing the inflow.
At a point $x_i, y_j$ in the square domain of size $L \times L$, the solution is given by a double integral over the area $[\min(0,x_i-L/2), \max(L,x_i+L/2)] \times [\min(0,y_j-L/2), \max(L,y_j+L/2)]$.
The portions of this area lying outside the domain (shaded grey and labelled B, C and D in Figure \ref{greenbc2dfig}) contribute to the boundary weighting.
A constant condition is imposed on each part of the inflow.
On the outflow boundaries (right and bottom) a zero Neumann condition is imposed.
The solution is given by
\begin{subeqnarray}\label{green2d3}
  u_{ij}^{n+1} &= M_{ijkl} (u_{kl}^n+\Delta t f_{kl}) \nonumber \\
  &+ B^L_{ij} g^L  + B^T_{ij} g^T + B^R_{ij} \cdot g^R_{ij} + B^B_{ij} \cdot g^B_{ij}, \\
  B^L_{ij} &= \sum_{k=1}^N \sum_{l=2j}^N M_{ijkl} + \frac{1}{2} \sum_{k=2i}^N \sum_{l=2j}^N M_{ijkl}, \slabel{bl} \\
  B^T_{ij} &= \sum_{k=2i}^N \sum_{l=1}^N M_{ijkl} + \frac{1}{2} \sum_{k=2i}^N \sum_{l=2j}^N M_{ijkl}, \slabel{bt} \\
  B^R_{ij} &= E_{ij} B^L_{ij}, \\
  B^B_{ij} &= E_{ij} B^T_{ij},
\end{subeqnarray}
where $E_{ij}$ is the exchange or reversal matrix of size $P^2$.
The first term in \eqref{bl} corresponds to the area labelled B in Figure \ref{greenbc2dfig}.
The first term in \eqref{bt} corresponds to C.
The second terms in \eqref{bl} and \eqref{bt} together represent D as an average of the contributions from the left and top boundaries.

The outflow boundaries are more complicated.
A zero Neumann condition is imposed on the right by extending the solution on the line $x=L$ - which may not be constant - horizontally to the right.
Likewise the solution on the line $y=0$ is extended vertically downwards.
The outflow boundary values are therefore vectors $g^R_{ij}$ and $g^B_{ij}$ and they are pointwise-multiplied by the outflow boundary influence vectors $B^R_{ij}$ and $B^B_{ij}$ respectively.

%%%%%%%%%%%%%%%%%%%%%%%%%%%%%%%%%%%%%%%%%%%%

\section{Low-rank approximation}

The solver multiplies a dense $N \times N$ matrix $M$ by the RHS so it has a complexity of $N^2$.
This can be dramatically reduced by computing a low-rank approximation $M_L$ of the matrix $M$.
Of particular interest are hierarchical algorithms such as the Fast Multipole Method (FMM) \cite{greengard:87} and $\mathcal H^2$ matrix framework \cite{hackbusch:2000} which reduce the complexity to $\mathcal O(N \log(N))$ or even $\mathcal O(N)$ and scale very well on modern architectures.
FMM and $\mathcal H^2$ matrices are typically applied to problems with less compact Green's functions but there is still scope to compress the Gaussian kernel by other hierarchical methods like the Fast Gauss Transform (FGT) \cite{greengard:91}; this is a topic for future papers.
As a first step, a simple non-hierarchical technique is employed: retain matrix entries $m_{ij}$ corresponding to `strong' interactions between the points $i$ and $j$, defined by some cutoff criterion.
All the other entries correspond to `weak' interactions and are zeroed.
Two choices of cutoff criterion are now presented in 1D and 2D.

\subsection{Simple low-rank approximations}

Two methods for rank reduction were chosen: by thresholding the matrix entries and by specifying the matrix bandwidth.
In the first method a minimum value is specified, below which the matrix entry is set to zero. In the 1D case the sparse matrix $M^L$ is constructed according to
\begin{equation}
  m^{L}_{ij} =
  \begin{cases}
     m_{ij}, & m_{ij} \geq t, \\
     0, & \text{otherwise}.
  \end{cases}
\end{equation}
The sparse matrices $M^{L1}$ and $M^{L2}$ used in the numerical tests are constructed with $t=1/(10N)$ and $t=1.e-5$ respectively.

In the second method, let $B < N/2$ be the low-rank matrix bandwidth.
Then the entries of the sparse matrix $M^L$ are given by
\begin{equation}
  m^{L}_{ij} =
  \begin{cases}
     m_{ij}, & i-B \leq j \leq i+B, \\
     0, & \text{otherwise},
  \end{cases}
\end{equation}
with the same periodic indexing as the full-rank matrix.
The sparse matrices $M^{L3}$ and $M^{L4}$ used in the numerical tests are constructed with $B = \text{round}(\log(N^2))$ and $B=13$ respectively.
The values of $t$ and $B$ were chosen to get a good balance of accuracy and cost at all resolutions in the tests in Section \ref{results}.

A similar approach was taken for the 2D low-rank solvers.
The thresholding and specified-bandwidth methods were applied.
Matrix $M^{L1}$ is constructed according to
\begin{equation}
  m^{L1}_{ijkl} =
  \begin{cases}
     m_{ijkl}, & m_{ijkl} > 1/(10N^2), \\
     0, & \text{otherwise}.
  \end{cases}
\end{equation}
Matrix $M^{L2}$ had a constant threshold value of $1e-5$.
The entries of specified-bandwidth matrix $M^{L3}$ are given by
\begin{equation}
  m^{L3}_{ijkl} =
  \begin{cases}
     m_{ijkl}, & (k-i)^2+(l-j)^2 \leq B^2, \ B = \text{round}(\log(N^2)), \\
     0, & \text{otherwise}.
  \end{cases}
\end{equation}
Similarly, matrix $M^{L4}$ is constructed with a constant bandwidth of $B=10$.

%%%%%%%%%%%%%%%%%%%%%%%%%%%%%%%%%%%%%%%%%

\section{1D numerical tests}\label{results}

The periodic results presented in this section have been published previously in \cite{bull:16b} while the non-periodic and 2D results are new.
The full-rank and low-rank direct solvers were implemented in Matlab.
For comparison a standard solution technique was also employed, namely the Matlab backslash operator for the solution of the sparse linear system \eqref{advdiff4}.
Since the matrices are Hermitian, the backslash operator defaults to a Cholesky decomposition or LDL factorisation.
A Gaussian initial condition was specified that decayed to zero within the domain.
In the non-periodic case, a zero Dirichlet boundary condition was specified on the left (inflow) and a zero Neumann condition on the right (outflow).
The advective and diffusive constants were set to $a=1.0$ and $\nu=0.05$ and simulations were run for 100 timesteps.
The domain length was $L=1$ and the following resolutions were used: $N=\{100, 200, 400, 800, 1600, 3200, 6400\}$.
For each value of $N$, the timestep was optimised for accuracy.
Further details of the optimisation are provided in \cite{bull:16b}.

\subsection{Error convergence}

\begin{figure}[hbtp]
  \subfigure[periodic]{\includegraphics[width=0.48\textwidth]{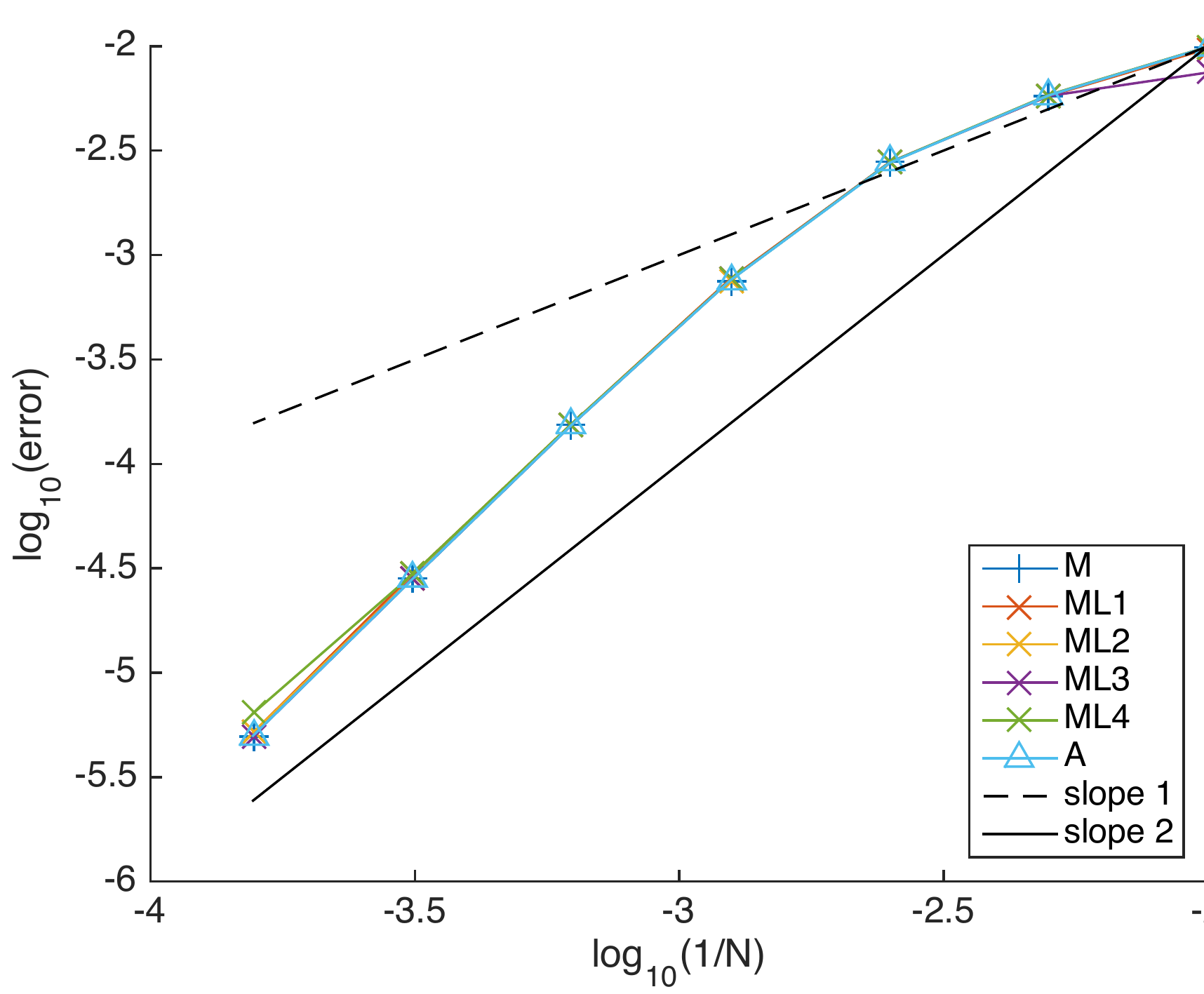}}
  \subfigure[non-periodic]{\includegraphics[width=0.48\textwidth]{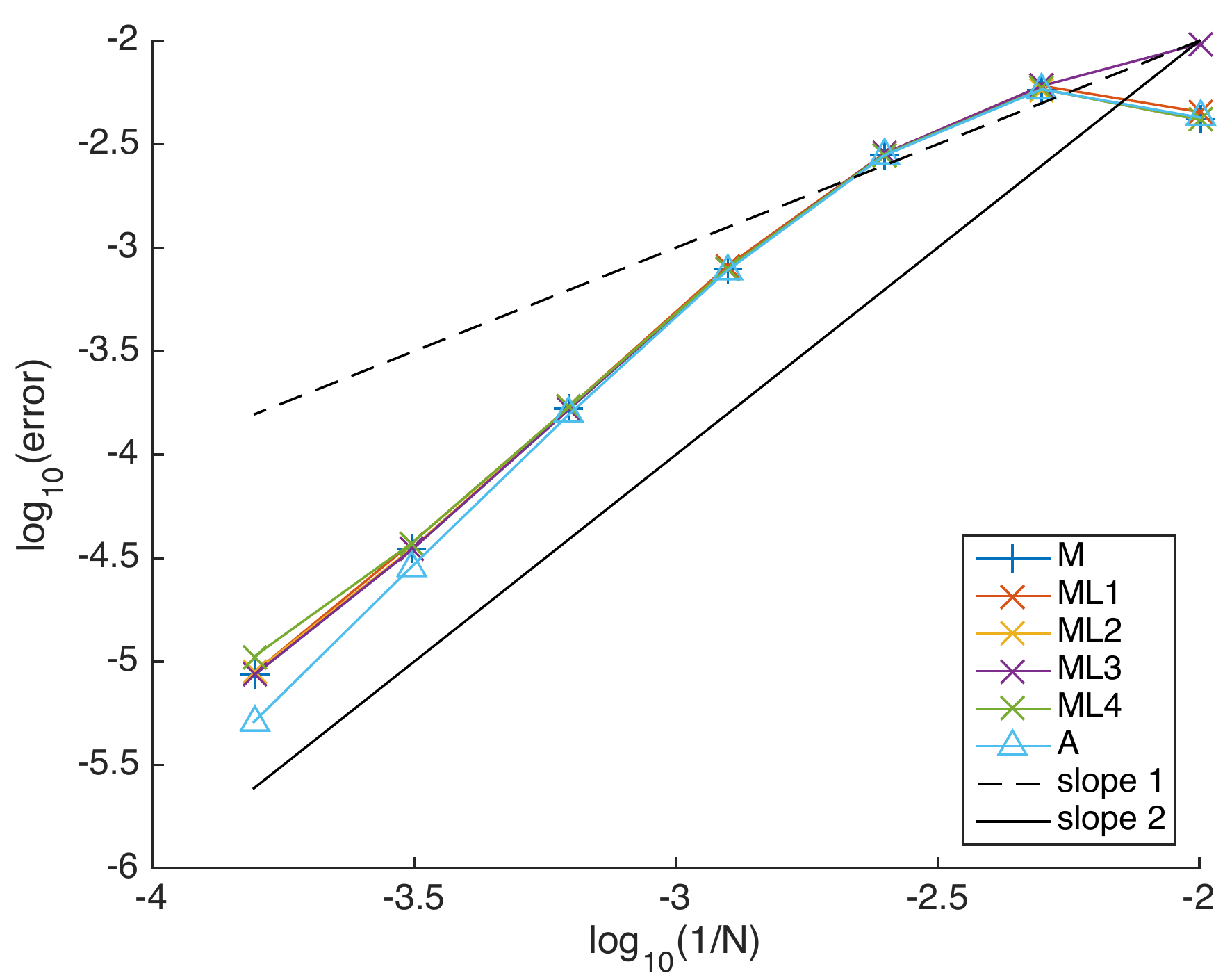}}
  \caption{Error convergence of 1D solutions. M = full-rank direct solver, A = backslash solver, $M^{L1-4}$ = low-rank solvers.}
  \label{err1d}
\end{figure}

The $L_2$ norm of the solution errors with respect to the exact solution was calculated.
Periodic and non-periodic results are plotted in Figure \ref{err1d}.
All the solvers achieve second-order error convergence.

\subsection{CPU time}

\begin{figure}[hbtp]
  \subfigure[periodic]{
  \includegraphics[width=0.48\textwidth]{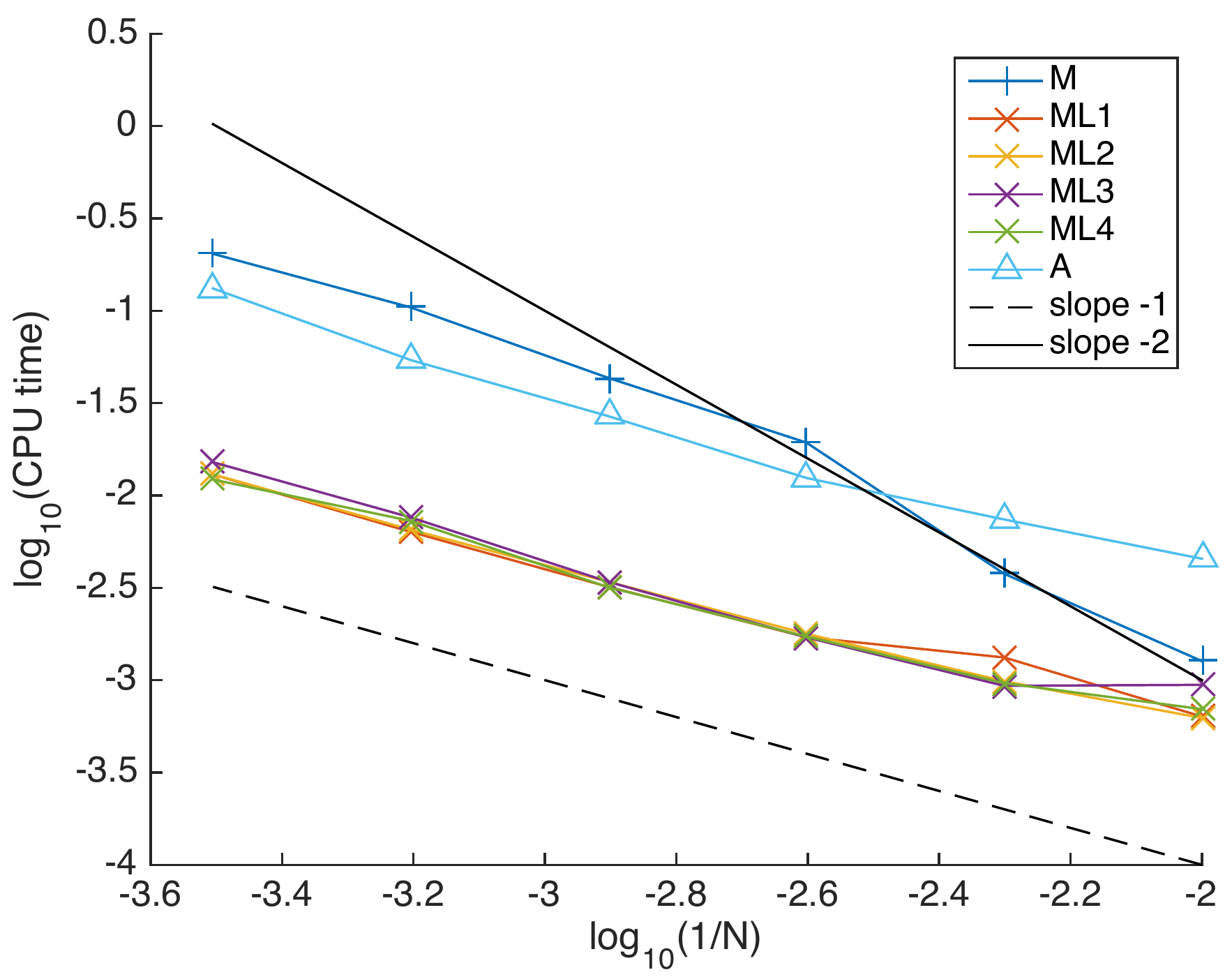}}
  \subfigure[non-periodic]{
  \includegraphics[width=0.48\textwidth]{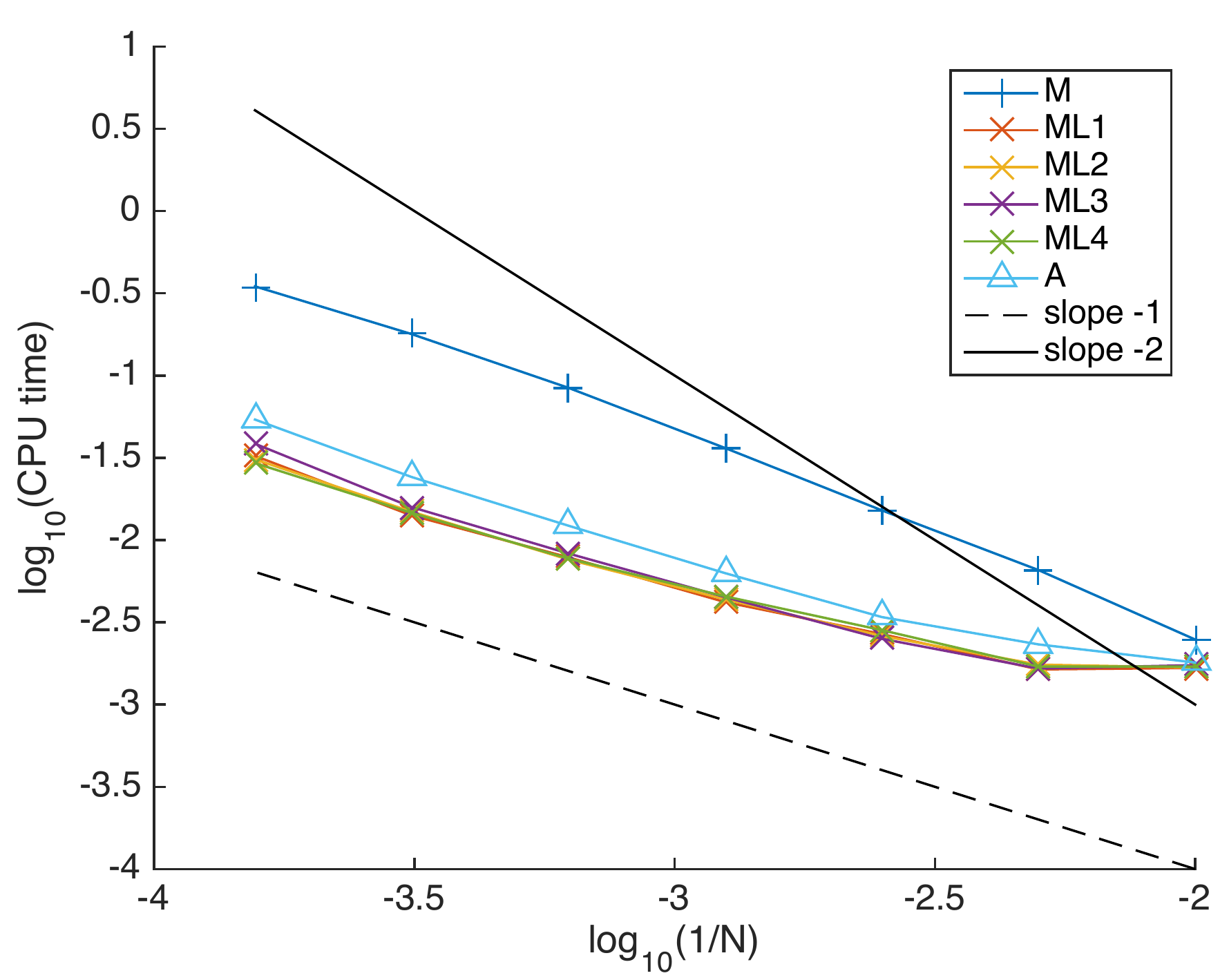}}
  \caption{CPU time of 1D solutions.}
  \label{cpu1d}
\end{figure}

Figure \ref{cpu1d} shows the CPU times for 100 timesteps excluding matrix assembly.
The reference solver is much faster in the non-periodic test.
Matlab's backslash operator automatically chooses the best solver for the matrix properties and it looks like a different solver has been chosen for each test.
The full-rank direct solver is the slowest in both tests by an order of magnitude relative to the low-rank solver, but it still obtains roughly linear scaling with $N$.
All the low-rank solvers obtain linear scaling and are roughly the same speed.

%%%%%%%%%%%%%%%%%%%%%%%%%%%%%%%%%%%%%%%%%

\section{2D numerical tests}\label{res2d}

In the 2D tests, the domain was $x,y=[0:1],[0:1]$ and the initial condition was a Gaussian distribution centred at $(0.5,0.5)$ and shifted vertically by $-0.5$:
\begin{equation}\label{ic2d}
  u_0(x,y) = -0.5+\exp(((x-0.5)^2+(y-0.5)^2)/\nu).
\end{equation}
The diffusion coefficient was $\nu = 0.01$ and the advection coefficient was $a=1$ with flow direction $\pi/4$ below the horizontal (toward bottom right corner).
\eqref{ic2d} decays to approximately zero within the domain, allowing for the imposition of homogeneous Dirichlet inlet conditions.
Five resolutions were used: $N=\{25 \times 25,50 \times 50,75 \times 75,100 \times 100,125 \times 125\}$.
The simulations proceeded for 25 timesteps with advective CFL number $c_A=1$.
The exact solution is
\begin{equation}
  u(x,y,t) = -0.5+\frac{1}{4t+1} \exp(((x-0.5)^2+(y-0.5)^2)/\nu (4t+1)).
\end{equation}
In the first test the boundaries were periodic.
In the second test the Dirichlet boundary condition $u=-0.5$ was imposed on the left and top, and zero Neumann boundaries on the right and bottom.

\subsection{Error convergence}

\begin{figure}[hbtp]
  \subfigure[periodic]{
  \includegraphics[width=0.48\textwidth]{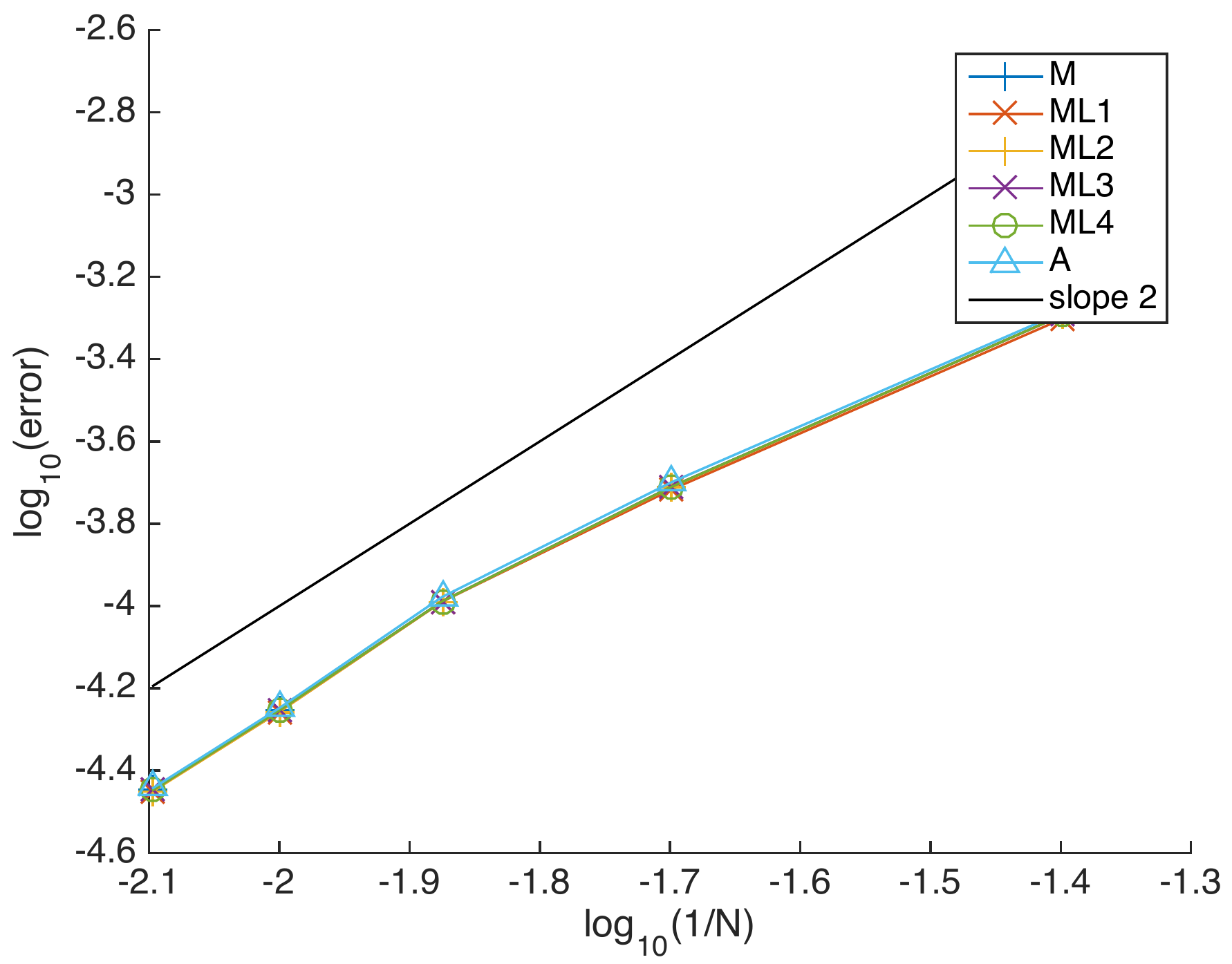}}
  \subfigure[non-periodic]{
  \includegraphics[width=0.48\textwidth]{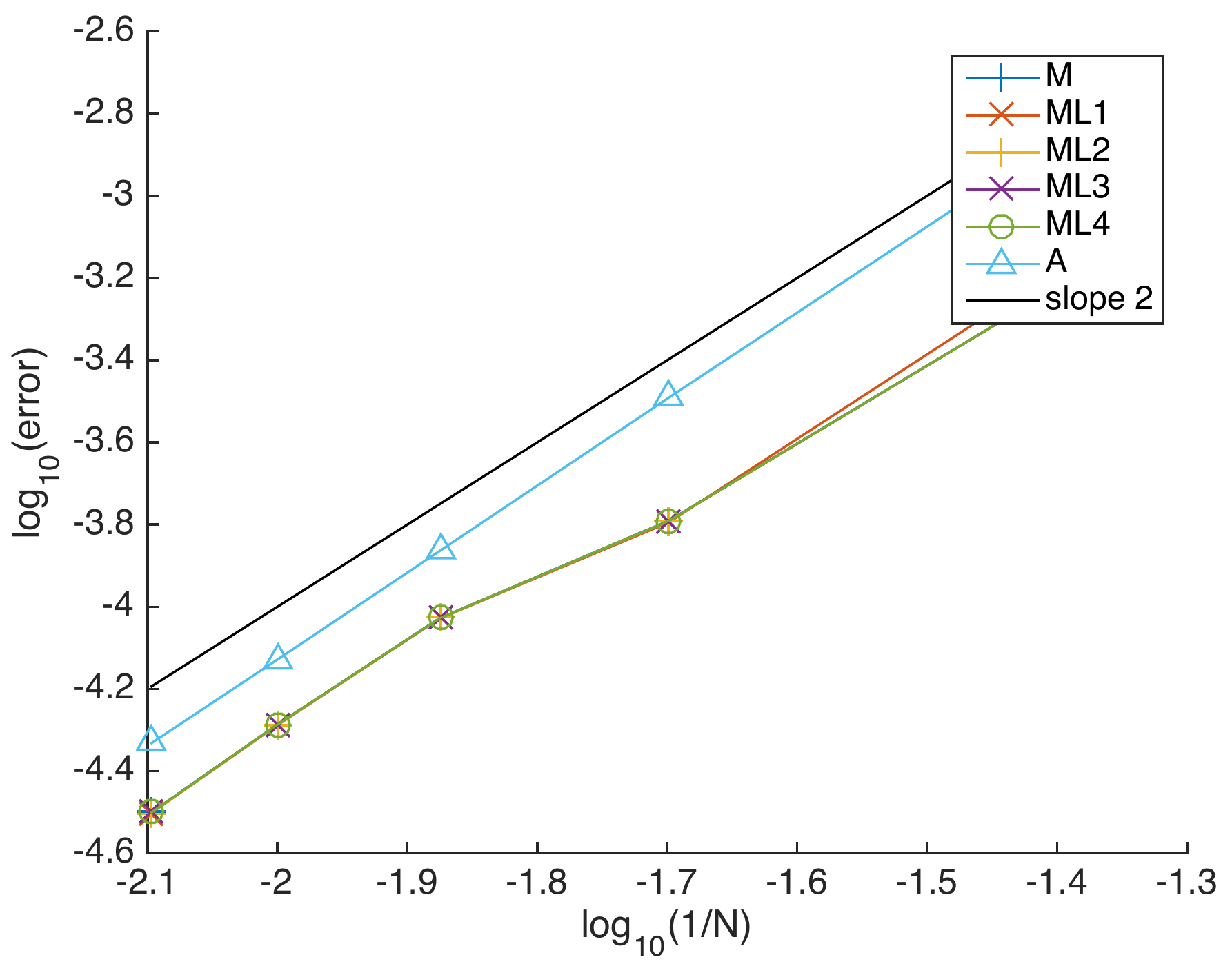}}
  \caption{Error convergence of 2D solutions.}
  \label{err2d}
\end{figure}

Figure \ref{err2d} plots the $L_2$ norm of the 2D periodic and non-periodic solution errors with respect to the analytical solution.
All schemes display $\mathcal O(N^2)$ error convergence.

\subsection{CPU time}

\begin{figure}[hbtp]
  \subfigure[periodic]{
  \includegraphics[width=0.48\textwidth]{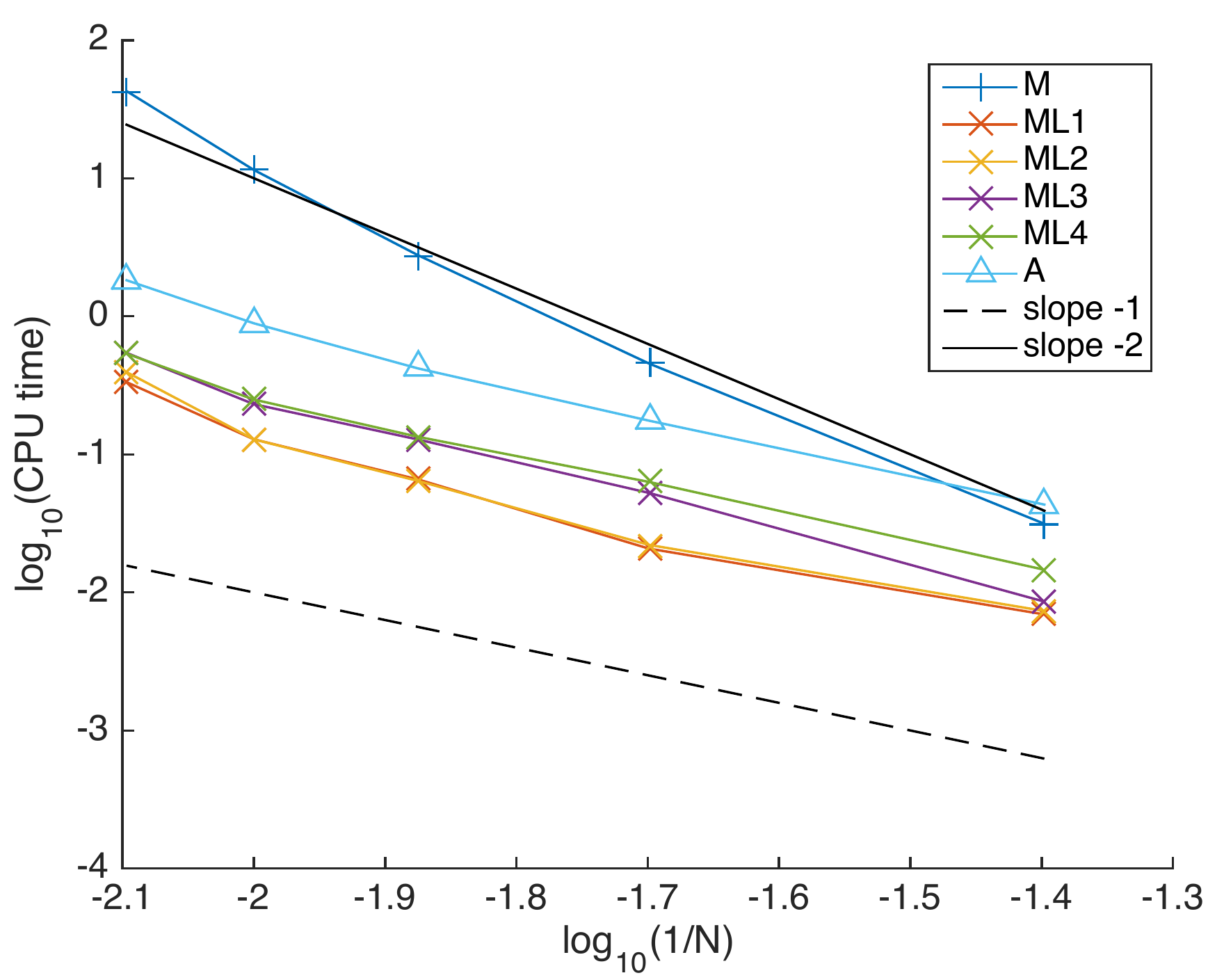}}
  \subfigure[non-periodic]{
  \includegraphics[width=0.48\textwidth]{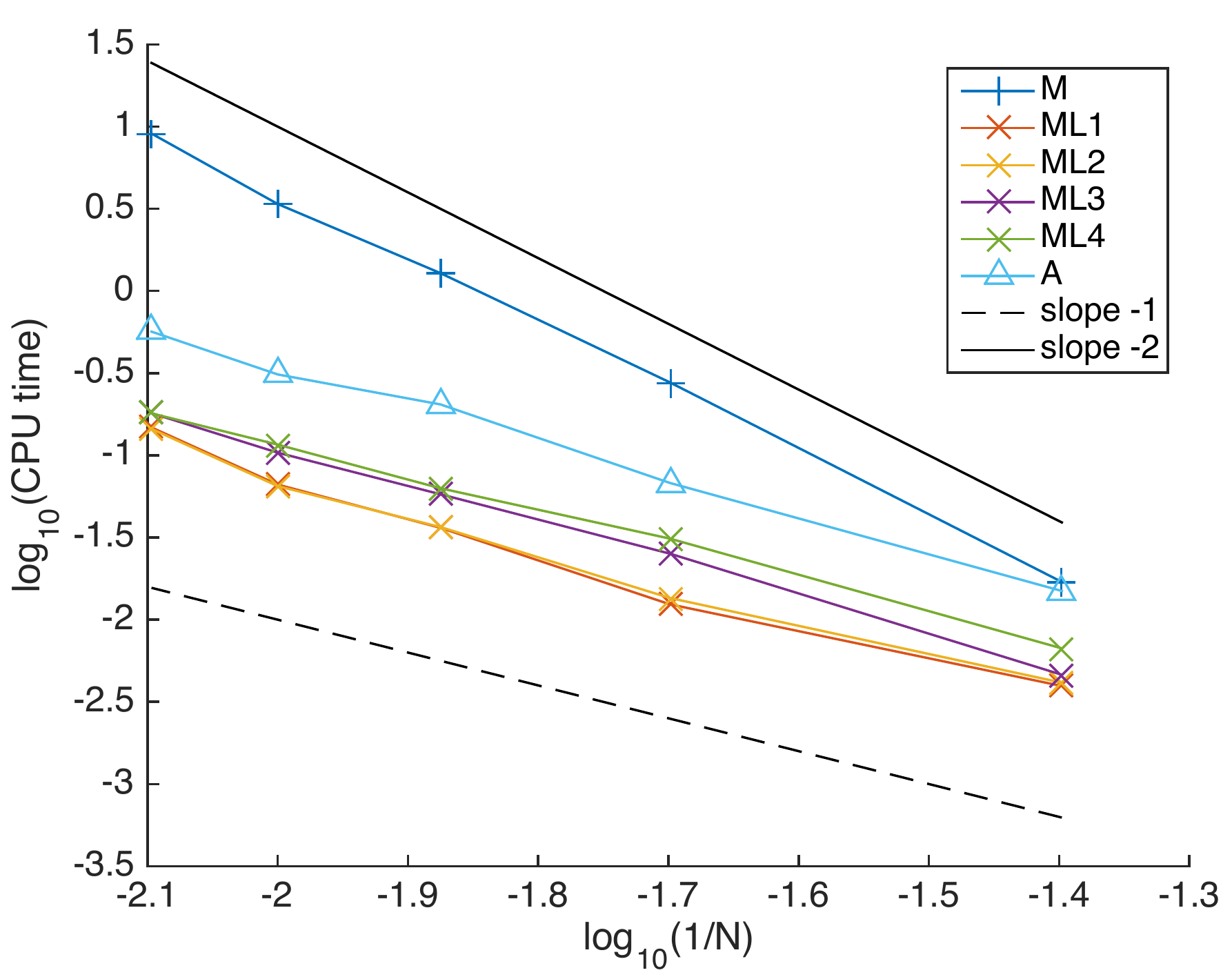}}
  \caption{CPU time of 2D solutions.}
  \label{cpu2d}
\end{figure}

Figure \ref{cpu2d} shows the CPU times for 25 timesteps excluding matrix assembly.
The full-rank solver requires approximately $\mathcal O(N^2)$ effort while the low-rank and reference solvers require only $\mathcal O(N)$ effort.
The low-rank direct solvers $M^{L1}$ and $M^{L2}$ are almost an order of magnitude faster than the reference solver.
All the low-rank solvers are faster even though they suffer no accuracy penalty compared to the full-rank and reference solvers.

\section{Conclusions}

Low-rank approximation of the discrete Green's function is a potentially powerful technique for solving or preconditioning large-scale stiff systems of elliptic and parabolic equations.
This paper outlines a simple approach for extending the method to hyperbolic problems.
The key step was to treat the hyperbolic advective term as a source, thereby transforming the equation into a parabolic form with a compact symmetric Green's function.
Thereafter a low-rank approximation of the discretised Green's function was calculated, leading to a second-order accurate direct solver with a computational cost proportional to $N$.

1D and 2D numerical tests on a serial processor validated the method.
Dirichlet and Neumann boundary conditions were imposed by the method of images.
The 2D numerical tests had a Reynolds number of 100 so the flow can be called advection-dominated.
Four different non-hierarchical low-rank approximations were tested in 1D and 2D.
All four demonstrated no loss of accuracy relative to the full-rank solver ($M$) but took roughly an order of magnitude less CPU time.
Moreover, all low-rank solvers obtained linear scaling of the CPU time with $N$.
Some tuning of the cutoff criteria was required to obtain best accuracy and cost.

Several improvements to the current method are under development including higher-order approximations of the time integral in \eqref{green1} and treatment of general boundary conditions and geometries.
Analysis and comparison of different HRLA techniques for the Gaussian kernel is also in progress.
The goal is to develop a fast solver based on HRLA for 2D and 3D stiff hyperbolic problems which could greatly reduce the computational effort required for very large $N$.
Of particular interest is to develop a massively parallelised solver/preconditioner for high-order accurate simulations of high-Reynolds number turbulent flows.

\begin{acknowledgement}
The author wishes to thank his colleagues Stefan Engblom and Sverker Holmgren for useful discussions and gratefully acknowledges financial support from the Liljewalch and Esseen travel scholarships.
\end{acknowledgement}

\bibliographystyle{elsarticle-num}

\end{document}